\documentclass[useAMS,usenatbib]{mn2e}
\usepackage{epsfig}

\newcommand{\alphaori}{$\alpha$ Ori~}
\newcommand{\Lsol}{L$_{\odot}$}
\newcommand{\Msol}{M$_{\odot}$}
\newcommand{\HI}{H\,{\sc {i}}~}
\newcommand{\CII}{C\,{\sc {ii}}~}

\newcommand{\Msold}{M$_{\odot}$\,yr$^{-1}$}
\newcommand{\Vstar}{V$_{\star}$}
\newcommand{\Vlsr}{V$_{\rm lsr}$}
\newcommand{\Vexp}{V$_{\rm exp}$}

\newcommand{\Teff}{T$_{\rm eff}$}
\newcommand{\kms}{km\,s$^{-1}$}

\newcommand{\lsim}{~\rlap{$<$}{\lower 1.0ex\hbox{$\sim$}}}
\newcommand{\gsim}{~\rlap{$>$}{\lower 1.0ex\hbox{$\sim$}}}

\title[The detached shell of \alphaori]{Discovery of a detached \HI gas shell 
surrounding $\alpha$ Orionis}
\author[T. Le Bertre, L. D. Matthews, E. G\'erard and Y. Libert]
{T. Le~Bertre$^{1}$, L. D. Matthews$^{2}$, E. G\'erard$^{3}$ and 
Y. Libert$^{4}$\\
$^{1}$LERMA, UMR\,8112, CNRS \& Observatoire de Paris, 61 av. de 
           l'Observatoire, F-75014 Paris, France\\
$^{2}$MIT Haystack Observatory, Off Route 40, Westford, MA 01886, USA\\
$^{3}$GEPI, UMR\,8111, CNRS \& Observatoire de Paris, 5 place J. Janssen, 
           F-92195 Meudon Cedex, France\\
$^{4}$IRAM, 300 rue de la Piscine, Domaine Universitaire, F-38406 Saint Martin 
d'H\`eres, France\\}
\begin{document}

\date{Accepted 2012 March 1.  Received 2012 February 24; in original form 2011 December 28}

\pagerange{\pageref{firstpage}--\pageref{lastpage}} \pubyear{2002}

\maketitle

\label{firstpage}

\begin{abstract}
We report the detection of the \HI line at 21 cm in the direction of \alphaori 
with the Nan\c cay Radiotelescope and with the Very Large Array. 

The observations confirm the previous detection of \HI emission centered on 
$\alpha$~Ori, but additionally reveal for the first time a quasi-stationary 
detached shell of neutral atomic hydrogen $\sim4'$ in diameter (0.24~pc 
at a distance of 200~pc). The detached shell appears elongated in a direction 
opposite to the star's space motion.

A simple model shows that this detached atomic gas shell can result from 
the collision of the stellar wind from \alphaori with the local interstellar 
medium (ISM). It implies that \alphaori has been losing matter at a rate of 
$\sim$ 1.2\,$\times$\,10$^{-6}$\,\Msold ~for the past 8\,$\times$\,10$^4$ 
years. 

In addition, we report the detection of atomic hydrogen associated with the 
far-infrared arc located 6$'$ north-east of $\alpha$ Ori, that has been 
suggested to trace the bow shock resulting from the motion of the star through 
the ISM. We report also the detection by the Galaxy Evolution Explorer 
(GALEX) of a far-UV counterpart to this arc. 

\end{abstract}


\section{Introduction}
Evolved late-type stars (AGB stars and red supergiants) are observed to 
undergo mass loss at high rates. They are thus surrounded by expanding 
circumstellar envelopes (CSEs). Observations show a wide variety of different 
properties of the central stars and of the CSEs, which are not always easily 
related. The time variability of the mass loss phenomenon may explain why it 
is difficult to relate the properties of CSEs to those of the central stars. 
Also, some stars may share the same observational properties, but be in 
slightly different evolutionary stages, with different initial masses and 
different mass loss histories. 
It would be extremely useful to get overviews of CSEs that would allow us 
to smooth the chaotic effects of the mass loss time-variability and as well 
to measure the total mass expelled by the central stars. With this goal 
in mind, we are exploring the properties of CSEs around evolved stars 
by using the atomic hydrogen (H\,{\sc {i}}) line at 21 cm. 

Indeed, hydrogen is expected to be the dominant species in stellar winds of 
evolved stars. It is also expected to be in atomic form right from the stellar 
atmosphere if the effective temperature, \Teff, is larger than 2500\,K,  
and to remain atomic in the expanding wind (Glassgold \& Huggins 1983). 
Also, atomic hydrogen should be protected from photoionization 
by the abundant atomic hydrogen in the interstellar medium (ISM) which absorbs 
UV photons of energy larger than 13.6 eV. The \HI line at 21 cm should thus be 
an important tracer of stellar winds for late-type stars 
with \Teff $>$ 2500\,K, including the outer CSEs where molecules 
are dissociated. \HI studies should bring unique information on the kinematics 
in these media, and in particular in the regions of interaction with the ISM. 

However the detection of red giants in the \HI line 
at 21 cm is hampered by the competing emission from atomic hydrogen in 
the ISM (which ironically protects the circumstellar atomic hydrogen from 
destruction) along the same lines of sight. For a long time this difficulty 
precluded \HI observational studies of evolved stars. However, recently, 
taking advantage of upgraded instruments, we addressed this topic and obtained 
several detections of AGB stars (G\'erard \& Le\,Bertre 2006, Matthews \& Reid 
2007). In general, a narrow emission line with a quasi-gaussian profile (FWHM 
$\sim$ 3\,\kms) is observed close to the stellar radial velocity (\Vstar). 
 
The fate of massive stars is strongly dependent on their mass loss history, 
which is still poorly understood, in particular when they are red supergiants 
(van Loon 2010). Our previous successes encouraged us to extend \HI 
observations to this class of stars. 
Although suffering from strong galactic \HI emission, the supergiant \alphaori 
was detected by Bowers \& Knapp (1987) using the Very Large Array. 
At the position of the star, they found emission from the front and back sides 
of a CSE expanding at a velocity of 14 \kms. 

In this paper, we report new 21-cm observations of 
$\alpha$~Ori that we analyze in combination with the previous data 
from Bowers \& Knapp (1987), in order to revisit in \HI this otherwise 
well-documented red supergiant.

\begin{figure}
\centering
\epsfig{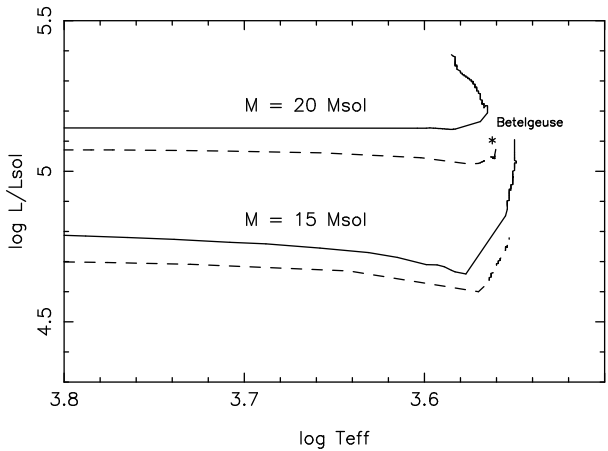}
 \caption{Evolutionary tracks from Meynet \& Maeder (2003; Z\,=\,0.02) 
for stars of initial masses 15 and 20 \Msol ~until the end of He-burning; 
full lines : models with rotation (V$_{\rm i}$ = 300 \kms), 
dashed lines: models without rotation. The star symbol (*) marks the position 
of \alphaori assuming a distance of 200\,pc (log L/\Lsol = 5.1; 
\Teff = 3650\,K).}
  \label{MeynetMaeder}
\end{figure}

\section[]{Basic data}
\label{basicdata}
\alphaori (Betelgeuse) is the prototype of red supergiants. It has a spectral 
type M2Iab. Perrin et al. (2004) have measured its effective temperature by 
infrared interferometry: \Teff=3641$\pm$53\,K. This determination fits well 
with the revised temperature scale of Galactic Red Supergiants by Levesque 
et al. (2005, 3650\,K for $\alpha$ Ori). It indicates clearly that atomic 
hydrogen should be the dominant species in the atmosphere and outflow of 
$\alpha$\,Ori.

Although it is the closest supergiant, its distance is not well known. 
Hipparcos measured a parallax of 7.63$\pm$1.64 (Perryman et al. 1997). 
It has been revised to 6.55$\pm$0.83 mas (van Leeuwen 2007), which 
translates to a distance of 153 pc (135-175 pc). But Harper et al. (2008), 
using multi-wavelength radio data obtained at the Very Large Array (VLA), 
derived a smaller parallax/larger distance. 
When combining the original Hipparcos results with VLA data, 
they get d=197$\pm$45\,pc. Harper et al. find also proper motions which are 
reduced with respect to Hipparcos. They suggest that the Hipparcos 
parallax is in error because of the large angular size of the star and 
the motions of the stellar photocenter due to convection 
(Lim et al. 1998, Haubois et al. 2009, Ohnaka et al. 2011). 
In the following we follow Harper et al. (2008) and adopt a nominal distance 
of 200~pc. At this distance, the bolometric luminosity corresponds to 
log L/\Lsol $\approx$ 5.10 $\pm$ 0.22, which, from the Meynet \& Maeder (2003) 
models (see Fig.~\ref{MeynetMaeder}), indicates a progenitor that was 
an O9V main-sequence star of 20~\Msol ~(Harper et al. 2008). Due to mass loss 
from the main sequence to the red supergiant stage, the present stellar mass 
would be $\sim$\,12--16\,\Msol. Recently using predictions of limb darkening 
by stellar atmosphere models and high spatial resolution observations at 
1.64\,$\mu$m, Neilson et al. (2011) estimate a present mass of 
11.6$^{+5.0}_{-3.9}$\Msol, in good agreement with Maeder \& Meynet's 
20\Msol-models. However, this should not be overemphasized since, as 
mentioned above, the mass loss history of red supergiants is rather insecure. 

The proper motions, revised by Harper et al. (2008), are 24.95 mas\,yr$^{-1}$ 
in right ascension, and 9.56 mas\,yr$^{-1}$ in declination. For a distance 
of 200 pc, and assuming a solar motion, U$_0$\,=\,11.1\,\kms, 
V$_0$\,=\,12.24\,\kms, and W$_0$\,=\,7.25\,\kms ~(Sch{\"o}nrich et al. 2010), 
we derive a velocity in the plane 
of the sky of 30\,\kms ~with an apex at PA=\,53$^{\circ}$.

\begin{figure}
\centering
\epsfig{figure=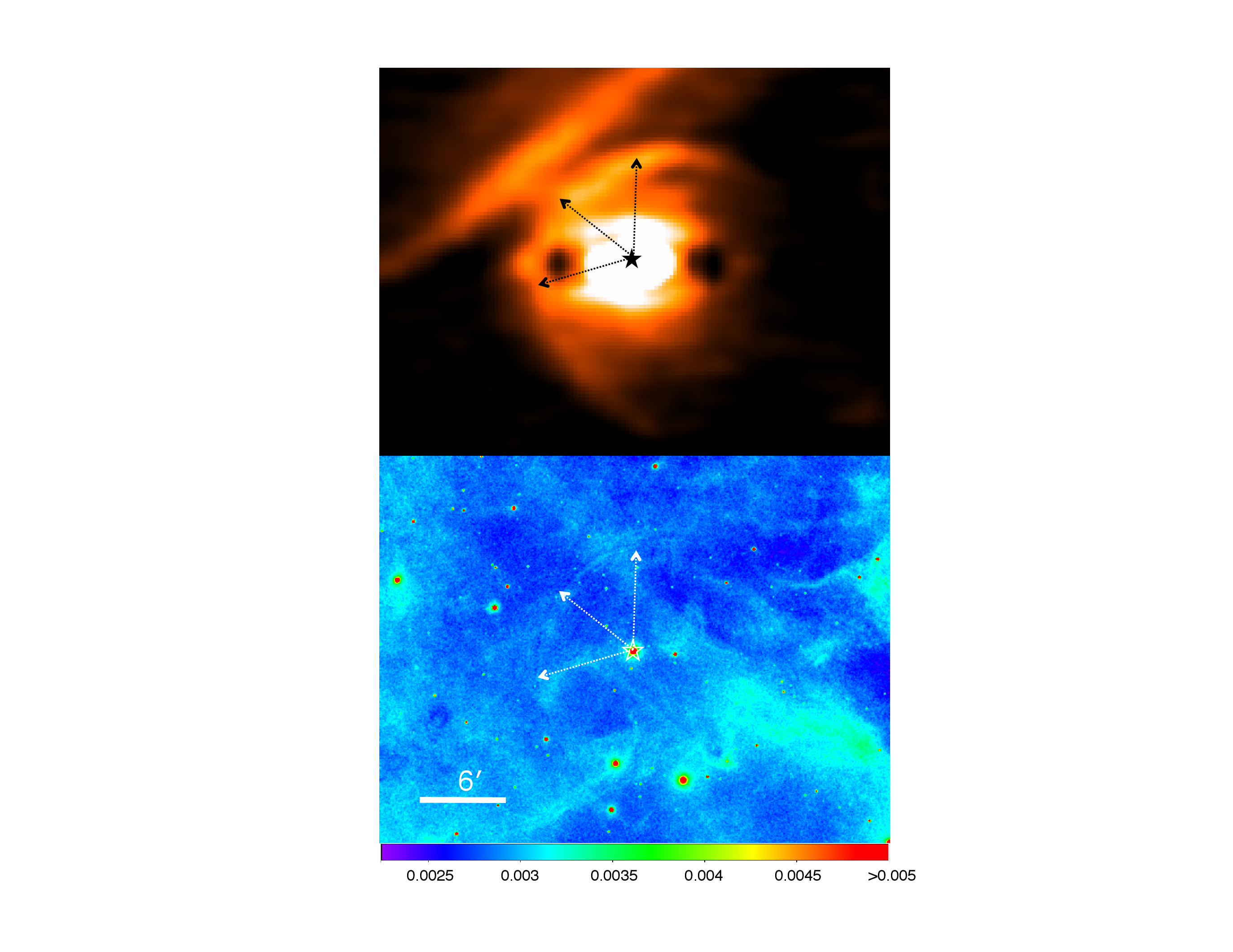,angle=-90,width=11.0cm}
 \caption{Upper panel: IRAS image at 60\,$\mu$m centered on $\alpha$ Ori 
(star symbol). Note the arc (which is pointed by the three arrows), 
$\sim$\,6$'$ north-east of $\alpha$ Ori, and the bar, 10$'$ north-east 
(Noriega-Crespo et al. 1997). 
Lower panel: the same field observed by GALEX. 
Note the faint narrow counterpart to the 60\,$\mu$m arc detected 
by IRAS from PA $\sim$\,0$^{\circ}$ to 100$^{\circ}$. The scale 
is in counts per second and per pixel (1.5$''$).}
  \label{alphaori_IRAS_GALEX}
\end{figure}

Like most supergiants, \alphaori is surrounded by an extended circumstellar 
envelope. The physical processes responsible for red supergiant mass-loss are 
unclear, but may be related to atmospheric convection (Lim et al. 1998, 
Ohnaka et al. 2011). The convective cells could levitate cool gas above 
the photosphere and lead to mass loss (Josselin \& Plez 2007, 
Chiavassa et al. 2011). Kervella et al. (2011) 
have obtained spectacular infrared images with the Very Large Telescope 
of the European Southern Observatory that show an inhomogeneous 
spatial distribution of the material ejected by the central star at scales 
corresponding to several stellar radii (0.5-3$''$). IRAS 
(Infrared Astronomical Satellite) detected an extended structure 
at 60 $\mu$m, which Young et al. (1993a) fitted with a detached shell 
model of inner radius 0.2$'$ (0.01 pc at 200 pc) and outer radius 10.5$'$ 
(0.6 pc). However, through a careful analysis of high-resolution IRAS data, 
Noriega-Crespo et al. (1997) discovered a ring ('far-IR arc') of mean radius 
6$'$, mostly confined north-east of \alphaori around PA= 60$\pm10^{\circ}$, 
in good agreement with the apex of the proper motion 
(Fig.~\ref{alphaori_IRAS_GALEX}). They interpret this structure as a bow shock 
resulting from the interaction of the supergiant wind with the ISM. 
Assuming that the IR emission is due to dust, Noriega-Crespo et al. evaluate 
the mass in the far-IR arc to be $\sim$ 0.034 \Msol ~(at 200 pc). This 
bow-shock structure was confirmed by Akari at far-infrared wavelengths (Ueta 
et al. 2008). A straight filament ('bar') is seen at 10$'$ from the central 
star ahead of the bow shock. Although perpendicular to the direction of motion 
and of maximum intensity close to the apex of the bow shock, it does not  
seem related to \alphaori in a simple manner, but may have contributed to 
the extended emission reported by Young et al. (1993a), which thus appears 
to be a blend of two distinct sources. 

Very recently, \alphaori was observed by Herschel at 70~$\mu$m and 160~$\mu$m 
(Cox et al. 2012). The images have an unprecedented spatial resolution 
(6$''$ and 12$''$, respectively) and bring many new details 
that could not be seen on IRAS and Akari images. 
In particular, they reveal sub-structures in the far-IR arc discovered 
by IRAS (upper panel of Fig.~\ref{alphaori_IRAS_GALEX}), which breaks up into  
at least three thin shells. 
It is also clear that these shells are not circular, and that 
the radii of curvature, close to apex, are larger than the distance 
to $\alpha$ Ori, as would be expected from a bow shock. 
There are also hints of additional arcs at smaller radii. 

Finally, our present analysis has uncovered a faint counterpart to this ring 
in far-ultraviolet (FUV $\sim$154 nm) imaging obtained by GALEX (Galaxy 
Evolution Explorer; Martin et al. 2005). 
The GALEX data presented in Fig.~\ref{alphaori_IRAS_GALEX} 
were obtained, from Nov. 29, 2008, to Jan. 14, 2009, 
as part of the Nearby Galaxies Survey 
(tile 5786) with an exposure time of 48915 sec. 
The GR6 pipeline calibrated data were retrieved from the GALEX archive 
(Morrissey et al. 2007), and smoothed by a Gaussian with kernel width 3 pixels 
(pixel\,$\equiv$\,1.5$''$). 
The FUV feature is coincident (in projection) with the outer shell 
revealed by Herschel. It can be seen most clearly north and northeast of 
the star, although there are hints that the entire structure subtends 
an angle of at least $\sim$105$^{\circ}$ 
(see arrows on Fig.~\ref{alphaori_IRAS_GALEX}). 
In the near-UV ($\sim$ 232 nm) the image is unfortunately saturated 
by a ghost of the central star. A similar, narrow and arc-shaped, UV-feature 
has been reported around IRC\,+10216 by Sahai \& Chronopoulos (2010). 
They ascribed the emission to collisionally excited H$_2$ molecules, although 
scattering of the interstellar radiation field by small dust particles 
could as well be considered. On the other hand, the \alphaori ~bar does not 
seem to be detected by GALEX. To our knowledge the far-UV emission associated 
with the far-IR arc of \alphaori has not been reported before. 

The circumstellar envelope of \alphaori has also been studied in the CO 
rotational lines. Huggins (1987) observed a CO(2-1) emission line centered at 
\Vstar=3.7$\pm$0.4\,\kms ~(LSR). It implies a 3D space velocity 
of 31\,\kms, mostly in the plane of the sky. 
From the profile, Huggins derives an expansion velocity, \Vexp=14.3\,\kms, 
and a mass-loss rate of $\sim$1$\times$10$^{-6}$ \Msold ~(scaled to 200 pc). 

\section[]{NRT single-dish observations}
\label{NRTobs}

The Nan\c cay Radiotelescope (NRT) is a meridian telescope with a rectangular 
aperture of effective 
dimensions 160\,m\,$\times$\,30\,m. The beam-width at 21 cm is 4$'$ in right 
ascension and 22$'$ in declination. The point source efficiency is 
1.4 K\,Jy$^{-1}$, and the system noise temperature, 35\,K. The observations 
of \alphaori have been obtained in the course of a large survey programme 
(G\'erard et al. 2011) over the period from October 2006 to September 2011.

\begin{figure}
	\centering
\epsfig{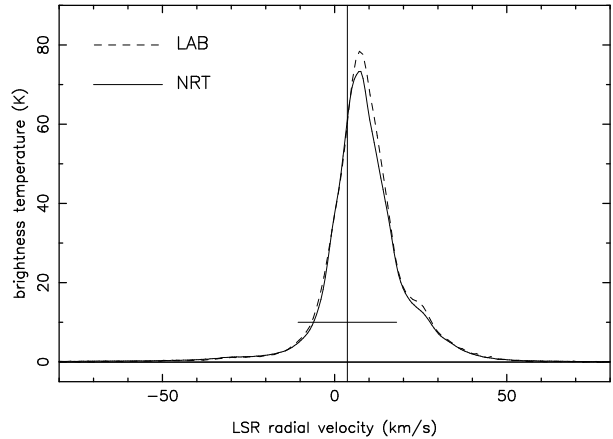}
 \caption{NRT frequency-switched spectrum obtained on the position of 
\alphaori (continuous line) compared to the LAB spectrum (dashed line, 
Kalberla et al. 2005). We adopted a main beam efficiency of 0.60 for the NRT 
at 21 cm. The vertical line marks the CO centroid velocity of \alphaori and 
the horizontal bar, the velocity extent of its CO emission (Huggins 1987).}
  \label{NRT-LAB}
\end{figure}

In Fig.~\ref{NRT-LAB}, we present a spectrum obtained 
in the frequency-switched 
mode of observation. It illustrates that in the direction of \alphaori the \HI 
emission is dominated by galactic emission due to interstellar atomic hydrogen 
along the same line-of-sight. The emission peaks at $\sim$ 7.5 \kms. 
Our spectrum is in good agreement with the spectrum extracted from 
the Leiden/Bonn/Argentina (LAB) survey (Kalberla et al. 2005; 
$\phi \sim 36'$). No distinctive feature can be seen in the velocity range 
expected from the CO observations of \alphaori (Huggins 1987).

\begin{figure*}
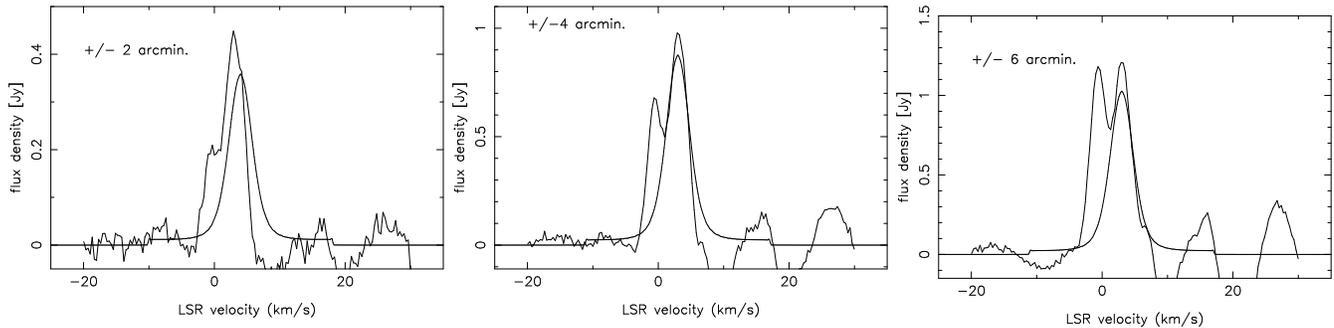
 
\centering
\epsfig{figure=fluxhan-C-0p5-NRT-mod19plusdata-PUb.ps,angle=270,width=5.8cm}
\epsfig{figure=fluxhan-C-1-NRT-mod19plusdata-PUb.ps,angle=270,width=5.8cm}
\epsfig{figure=fluxhan-C-1p5-NRT-mod19plusdata-PUb.ps,angle=270,width=5.8cm}
 \caption{\alphaori spectra obtained with the NRT in the position-switched 
mode. Left: off-positions taken at $\pm 2'$. Centre: off-positions taken at 
$\pm 4'$. Right: off-positions taken at $\pm 6'$. The thin lines correspond 
to the modeled spectra discussed in Sect.~\ref{interpretation}. }
  \label{NRTplusmodel}
\end{figure*}

Data were also obtained in the position-switched mode of observation with the 
central star placed at the ``on-position'', and the ``off-positions'' taken 
at $\pm 2'$, $\pm 4'$, and $\pm 6'$ in the east-west direction 
(Fig.~\ref{NRTplusmodel}). 
These spectra are strongly affected by interstellar confusion. 
Nevertheless, an emission feature is detected at the radial velocity expected 
for \alphaori ~(3 \kms). This is seen in the spectrum with the 
smallest throw ($2'$$\equiv$ 1/2 beam in right ascension, left panel) with 
a peak intensity of $\sim$ 0.5 Jy. The intensity is about 1\,Jy at $\pm 4'$ 
(1 beam, middle panel) and about 1.2 Jy at $\pm 6'$ (1.5 beam, right panel). 
However, the confusion increases rapidly with throw (see for instance 
the emission feature at --2\,\kms, and the absorption troughs at +10 
and +20 \kms, that may affect the \alphaori ~line-profiles). 
We have also obtained spectra with larger throws, but the confusion 
becomes so large that the emission feature at 3 \kms ~gets drowned.

Finally, position-switched spectra were obtained with the on-positions 
at +11$'$ (1/2 beam north) and at --11$'$ (1/2 beam south) of the central 
star. We find that the \HI source is centered on the stellar position in 
declination and that there is no evidence of an extension in the north-south 
direction (FWHM \lsim\,11$'$). 

Despite the high level of confusion, we can infer that an \HI source, 
coincident with $\alpha$\,Ori, is detected at \Vlsr = 3 \kms. This source 
does not appear extended in right ascension as compared to the NRT beam 
(diameter $\sim 4'$). The line profile is narrow with a FWHM $\sim$ 3 \kms.
The integrated intensity is $\approx$\,5\,Jy$\times$\kms, 
which translates to $\approx$\,0.05 \Msol ~in atomic hydrogen at 200 pc.
This relatively strong emission in \HI at a velocity for which Bowers \& Knapp 
(1987) did not report any signal was puzzling and motivated a reassessment 
of the case at the  Very Large Array (VLA). 

\section[]{VLA Observations} 
\label{VLAobs}

\subsection{C Configuration Observations}
\label{VLAobsC}
Bowers \& Knapp (1987) previously observed $\alpha$~Ori in the 
{\mbox{H\,{\sc i}}} 21-cm line with the VLA
of the National Radio Astronomy Observatory (NRAO) in its C configuration 
(0.08 to 3.2~km baselines).\footnote{The National
Radio Astronomy Observatory is operated by Associated Universities,
Inc., under cooperative agreement with the National Science Foundation.}  
These data comprise 32 frequency channels in a single (right circular) 
polarization with a channel spacing of 6.10~kHz (1.29~\kms) after on-line 
Hanning smoothing, resulting in a total bandwidth of 0.189~MHz 
($\sim$41~\kms). The band was centered at an LSR velocity of $+5.0$~\kms. 
In total, $\sim$5.5 hours of integration time was spent on-source. 
Further details on the observations can be found in Bowers \& Knapp (1987). 

We downloaded the Bowers \& Knapp data from the VLA archive and
reprocessed them, following standard procedures for data flagging and
gain calibration within the Astronomical Image Processing System (AIPS).  
The flux densities of our calibration sources are summarized in 
Table~\ref{VLA-calib}. 
No bandpass calibration was possible owing to contamination of the bandpass 
calibrator by {\mbox{H\,{\sc i}}} absorption across more than half of the band.

\subsection{D Configuration Observations}
\label{VLAobsD}
To obtain improved sensitivity to large-scale {\mbox{H\,{\sc i}}} emission 
toward $\alpha$~Ori, we obtained additional VLA observations of the star
using the VLA D configuration (0.035-1.0~km baselines) over the course of two
observing sessions on 2010 January 2-3 (4 hours) and 2010 January
3-4 (6 hours). In total, $\sim$6.7 hours were spent on-source.

The VLA correlator was used in 1A mode with a 1.56~MHz bandpass, yielding 
512 spectral channels with 3.05~kHz ($\sim$0.64~\kms) spacing in a single 
(right circular) polarization. The band was centered 
at an LSR velocity of $3.7$~\kms.
Observations of $\alpha$~Ori were interspersed with observations of a phase
calibrator, 0632+103, approximately every 20 minutes. 3C48 (0137+331) was used 
as a flux calibrator, and an additional strong point source (0521+166) 
was observed as a bandpass calibrator (Table~\ref{VLA-calib}). 
To insure that the absolute flux scale and bandpass calibration were not
corrupted by Galactic emission in the band, the flux and bandpass 
calibrators were each observed twice, with frequency shifts of $+$1.27~MHz 
and $-1.19$~MHz, respectively, relative to the band center used
for the observations of $\alpha$~Ori. 0632+103 was also observed once 
at each of these offset frequencies to permit more accurate bootstrapping 
of the absolute flux scale  to the $\alpha$~Ori data. We estimate that 
the resulting absolute flux scale has an uncertainty of $\sim$10\%.

%

\begin{table*}
  \caption{VLA Calibration Sources}
  \begin{tabular}{lccll}
  \hline

Source & $\alpha$(J2000.0) & $\delta$(J2000.0) & Flux Density (Jy) & Date\\
 \hline

3C48$^{\rm a}$ & 01 37 41.29 & +33 09 35.13 & 16.038$^{*}$  
& 1985 Sep 22\\

... & ... & ... & 15.8778$^{*}$   & 2010 Jan 2-4\\

0521+166$^{\rm b}$& 05 21 09.88 & 16 38 22.05  &
8.66$\pm$0.08$^{\dagger}$ & 2010 Jan 2-3\\

...  & ...  & ... & 
8.40$\pm$0.12$^{\dagger}$  & 2010 Jan 3-4\\

0532+075$^{\rm c}$ & 05 32 38.99 &  07 32 43.34 &
2.11$\pm$0.01 & 1985 Sep 22\\

0632+103$^{\rm c}$  & 06 32 15.32  &  10 22 01.67  &
2.39$\pm$0.05$^{\dagger}$ & 2010 Jan 2-3\\

...  & ...  & ... & 
2.25$\pm$0.04$^{\dagger}$ & 2010 Jan 3-4\\
\hline
\end{tabular}

Units of right ascension are hours, minutes, and
seconds, and units of declination are degrees, arcminutes, and
arcseconds. \\
{* Adopted flux density at 1420.76~MHz.}\\
{$\dagger$ Mean computed flux density from
observations at 1419.57 and 1422.0~MHz; see \S~\ref{VLAobs}.}\\

{a Primary flux calibrator}\\
{b Bandpass calibrator}\\
{c Phase calibrator}\\
\label{VLA-calib}
\end{table*}

Data processing was performed using AIPS. At the time of our observations, 
the VLA contained 21 operational antennas with L-band receivers, 
19 of which had been retrofitted as part of the Expanded Very Large Array 
(EVLA) upgrade. Data obtained during this EVLA transition period require 
special care during calibration.\footnote{See
http://www.vla.nrao.edu/astro/guides/evlareturn/.}
After applying the latest available corrections to the
antenna positions and performing an initial excision of corrupted data,
we  computed and applied a bandpass calibration to our
spectral line data to remove closure errors on VLA-EVLA baselines. 
The bandpass was normalized using channels 164-447, thus excluding 
the portion of the band affected by aliasing. We next computed
a frequency-averaged (channel~0) data set for use in calibrating
the frequency-independent complex gains, again using channels 164-447.
Following gain calibration, we applied time-dependent frequency
shifts to the data to compensate for changes caused by
the Earth's motion during the course of the observations. At this
stage, we also applied Hanning smoothing in velocity and
discarded every other channel, yielding a 256 channel data set with a
velocity resolution of $\sim$1.3~\kms. There were no continuum sources 
brighter than $\sim$0.02~Jy within our primary beam and none detected 
within a single spectral channel within $\sim10'$ of the stellar position, 
hence no continuum subtraction was performed.

\subsection{A Combined VLA Data Set}
\label{VLAobsCplusD}
To obtain a data set with the maximum possible $u$-$v$ coverage, we
have concatenated the VLA C and D configuration data over the velocity
range where the two data sets overlap ($-14.3\ge V_{\rm LSR}\ge
24.3$~\kms). This velocity range is sufficient to encompass the range
of CO velocities previously observed toward $\alpha$~Ori 
(see Fig.~\ref{NRT-LAB}). 
The $u$, $v$, and $w$ values of the C configuration data were precessed 
to J2000.0 values prior to concatenation.

We imaged the combined data using the AIPS IMAGR program with `robust +1' 
weighting of the visibilities. 
The large-scale Galactic emission along the line-of-sight 
(Fig.~\ref{NRT-LAB}) is poorly spatially sampled by the VLA, even
in the combined C+D configuration data set, resulting in patterns of
strong positive and negative mottling across a number of the channel images.  
Because such emission cannot be readily deconvolved by standard CLEAN
deconvolution algorithms, we performed only a shallow clean (20 iterations)
on each spectral channel to reduce sidelobes from strong compact emission
features. The rms noise per channel is variable 
and in most channels is dominated by the complex, large-scale emission. 

\begin{figure*}
 \epsfig{figure=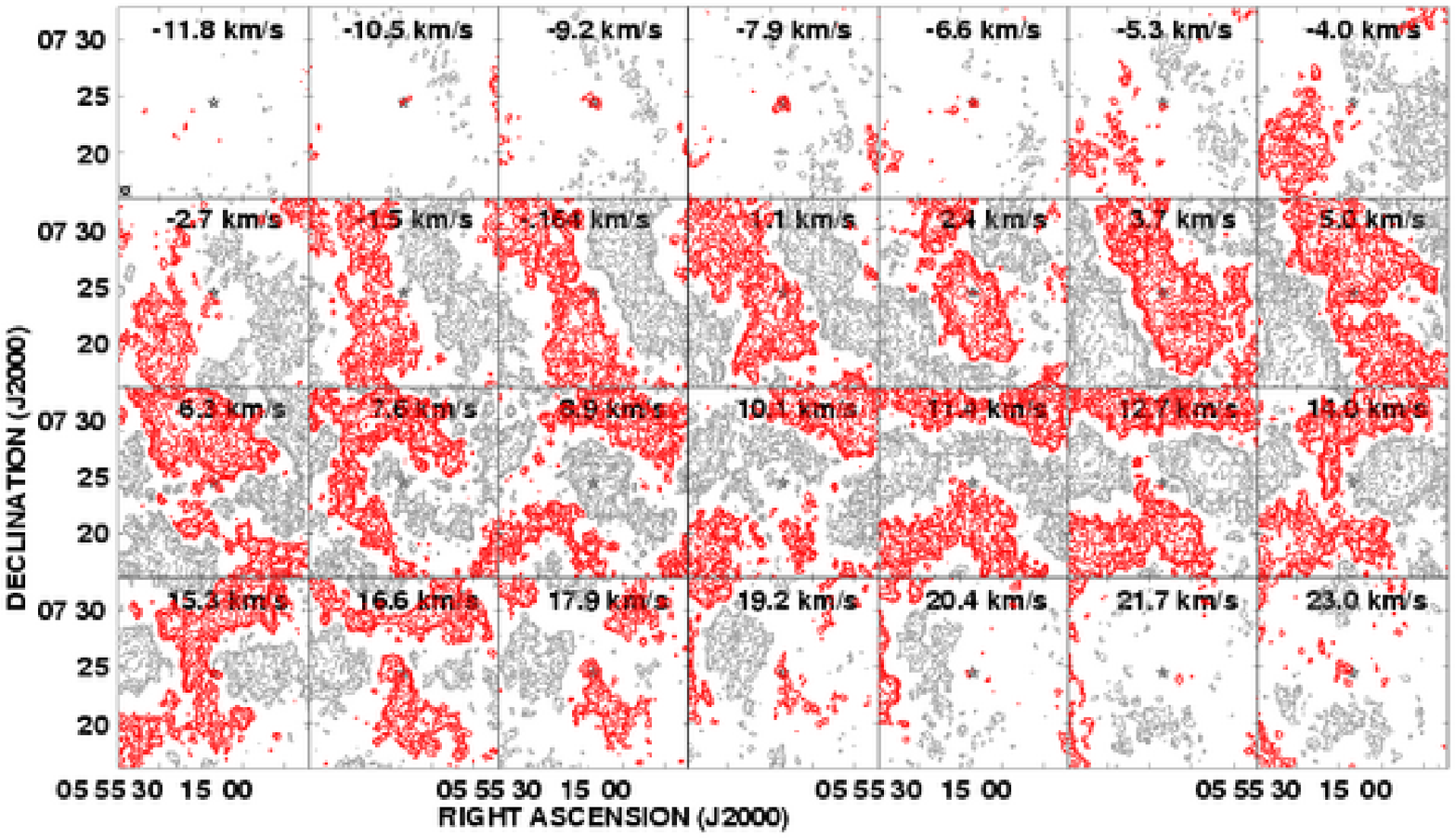,angle=270,width=18.0cm}
  \caption{Channel maps obtained at the VLA by combining the data in 
configuration C of Bowers \& Knapp (1987) with our 2010 data in configuration 
D. The data have a "robust +1" weighting, and the synthesized beam has 
a size of 37.4$''\times$32.0$''$. The contour levels are 
(-11.2, -8, -5.6, -4, -2.8, -2, 2..., 11.2)$\times$3.9 mJy/beam. 
The negative contours are in gray, and the positive contours in red. 
The star symbol marks the position of $\alpha$ Ori.}
  \label{VLA-channel-maps-full}
\end{figure*}

\begin{figure*}
\epsfig{figure=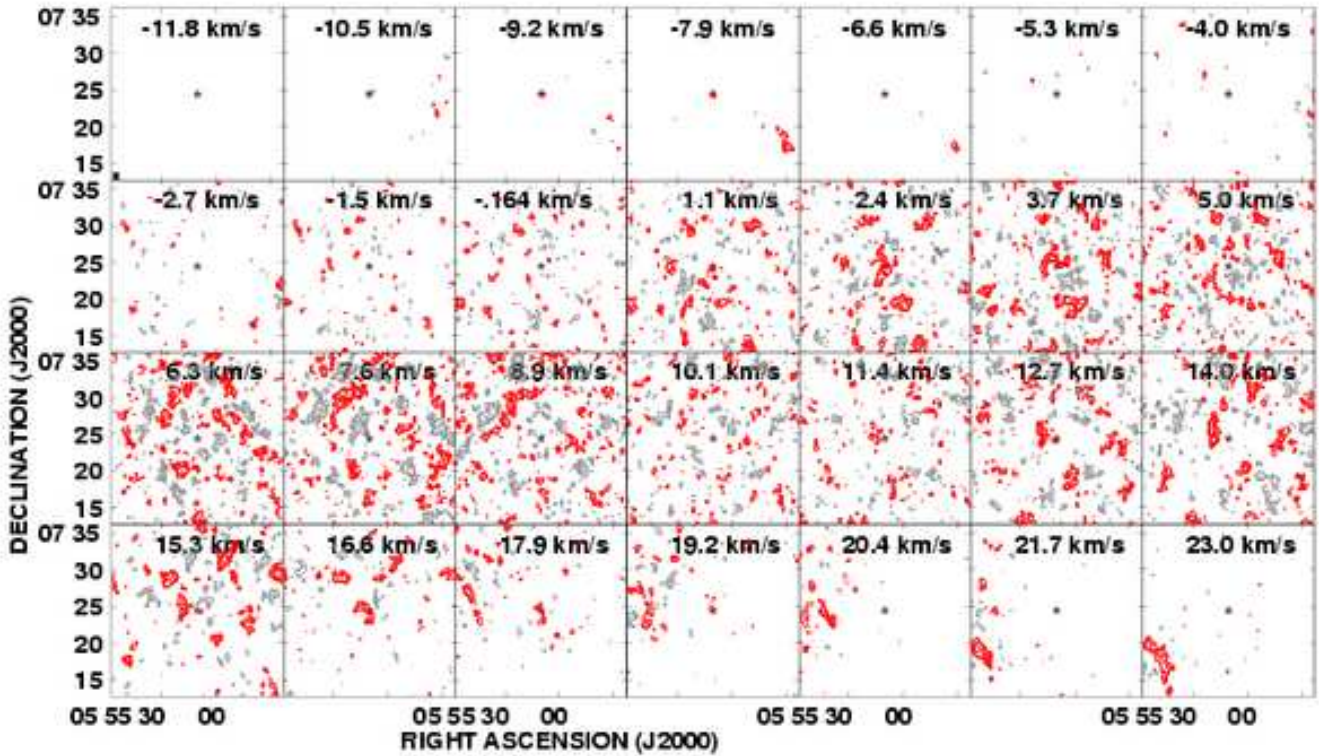,angle=270,width=18.0cm}
  \caption{Same as Fig.~\ref{VLA-channel-maps-full}, but selecting baselines 
larger than 0.4 k$\lambda$.}
  \label{VLA-channel-maps-HR}
\end{figure*}

\begin{figure}
\centering
\epsfig{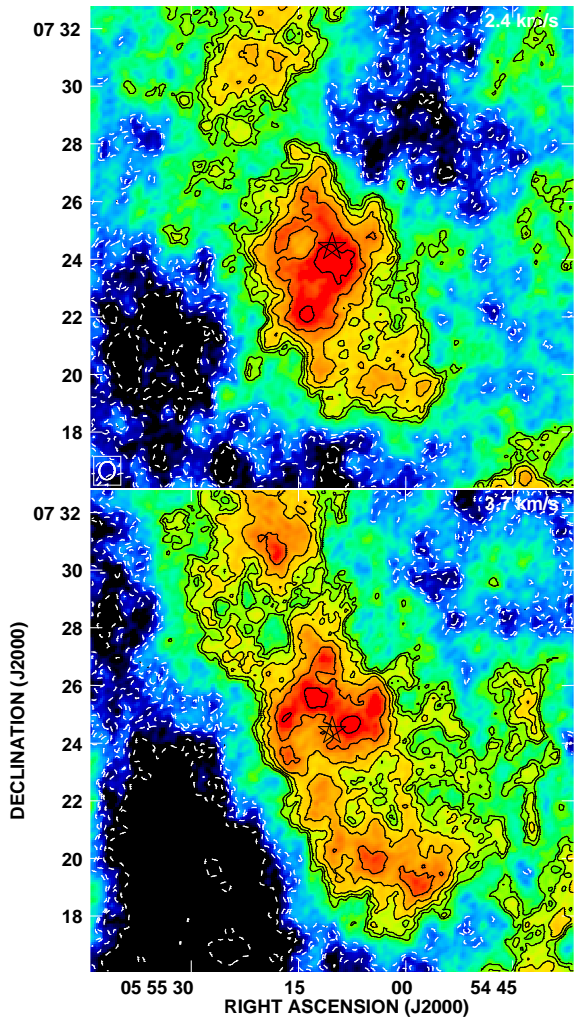}
\caption{Channel maps at 2.4\,\kms ~(top) and at 3.7\,\kms ~(bottom). 
The star symbol marks the position of $\alpha$ Ori. The contour levels are 
(-11.2, -8, -5.6, -4, -2, 2,..., 11.2)$\times$3.9 mJy/beam. 
Dashed lines represent negative contours. Note, in the map at 3.7 \kms, 
the emission peaks that surround the central star and appear to define 
the \alphaori gas detached shell.}
  \label{VLA-channel-map-3p1}
\end{figure}

\begin{figure}
\centering
\epsfig{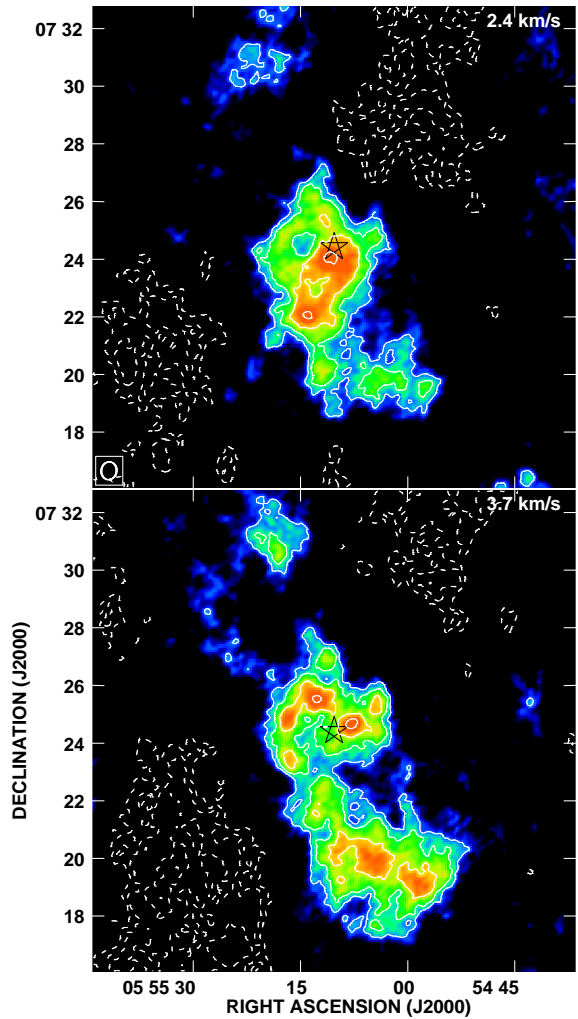}
\caption{Same as Fig.~\ref{VLA-channel-map-3p1}, but restricted to baselines 
larger than 0.2\,k$\lambda$. The contour levels here are 
(-8, -5.6, -4, 4, 5.6, 8, 11.2)$\times$3.9 mJy/beam.}
  \label{VLA-channel-map-3p1-restricted}
\end{figure}

Despite these complications, we find clear evidence of \HI emission associated 
with $\alpha$ Ori. The channel maps from --11.8 to +23.0~\kms ~are presented 
in Fig.~\ref{VLA-channel-maps-full}. They are strongly affected by extended 
interstellar emission for velocities above --5\,\kms. Nevertheless 
an isolated feature coincident with \alphaori ~is clearly seen from $-10.5$  
to $-5.3$\,\kms. In order to facilitate the identification of small 
sources, we present in Fig.~\ref{VLA-channel-maps-HR} the channel maps 
obtained when selecting only the largest $u$-$v$ spacings 
($\geq$ 0.4\,k$\lambda$). The spatial resolution is $\sim$~34$''$. 
A source coincident with the central star is 
detected in the channel maps from --9.2 to --6.6 \kms, 
and from +17.9 to +19.2 \kms. These detections are obtained in channels 
which correspond approximately to the extreme velocities 
expected from the wind of \alphaori ~(\Vstar--\Vexp= $-$10.6\,\kms, 
\Vstar+\Vexp=\,18.0\,\kms). 

We find also emission associated with \alphaori ~at +2.4~\kms ~and 
+3.7~\kms. Enlarged versions of the two channel maps are shown in 
Fig.~\ref{VLA-channel-map-3p1} (all baselines) and in 
Fig.~\ref{VLA-channel-map-3p1-restricted} (limited to baselines larger 
than 0.2\,k$\lambda$). At +2.4 \kms ~(upper panels), the emission 
peaks on the central star. On the other hand, in the +3.7 \kms ~channel 
map, one notes rather a ring of emission peaks surrounding the position of 
the star (bottom panels in Figs.~\ref{VLA-channel-map-3p1} 
and \ref{VLA-channel-map-3p1-restricted}). Outside this ring 
of $\sim 4'$ in size the emission decreases steeply, except south-west 
where a plateau of emission seems to extend away from the star. 
The \HI emission associated with \alphaori ~in these two channels fits well 
with that observed at the NRT (Sect.~\ref{NRTobs}).

\begin{figure}
\centering
\epsfig{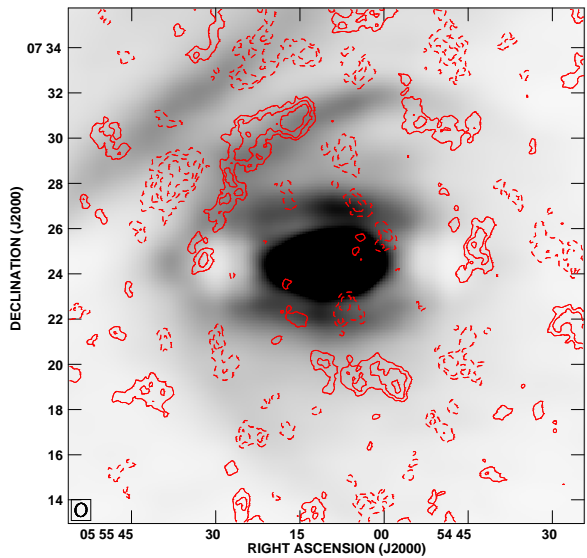}
\caption{\HI emission from Fig.~\ref{VLA-channel-maps-HR}, 
summed from -1.5 to 8.9 km/s. The contour levels are 
(-8., -5.6, -4., -2.8, 2.8, 4., 5.6, 8.)$\times$1.8 mJy/beam. 
The background image is from IRAS (see Fig.~\ref{alphaori_IRAS_GALEX}). 
Note the association between \HI emission and the far-IR arc.}
  \label{VLA-channel-maps-plus-IRAS}
\end{figure}

In addition, the channel maps from +6.3 to +10.1\,\kms 
~(Fig.~\ref{VLA-channel-maps-HR}) show \HI emission seemingly associated with 
the arc of far-infrared emission discovered by IRAS 6$'$ north-east of 
\alphaori ~(Noriega-Crespo et al. 1997). 
Several clumps are also noticeable along the far-IR arc at lower velocities, 
from --1.5\,\kms ~to +5\,\kms, and at +10\,\kms. 
In Fig.~\ref{VLA-channel-maps-plus-IRAS}, the \HI emission integrated 
over the range --1.5 to 8.9 \kms ~is overlaid on the IRAS image.  
The integrated intensity of the \HI emission associated with the far-IR arc 
is $\sim$\,4.9 Jy$\times$\kms. This is rather uncertain since the structures 
blend with background emission at low levels and because the maps in 
Fig.~\ref{VLA-channel-maps-HR} are probably missing some flux. 
It translates to a mass of $\sim$\,0.047\,\Msol ~in atomic hydrogen. 
Assuming 10\% in helium, the total mass can be estimated to be 
$\sim$\,0.068\,\Msol, which is within a factor 2 of the Noriega-Crespo et al. 
(1997) estimate for the mass in the far-IR arc. There is perhaps also 
\HI emission at +7.6\,\kms ~associated with the bar ahead of the arc, 
but the spatial coincidence is less convincing. 

Based on their C configuration data, Bowers \& Knapp (1987) reported emission 
associated with \alphaori ~at $\sim-$9\,\kms ~and at $\sim$+17\,\kms, but not 
at $\sim$+3 \kms. The reason is that they presented results based only on the 
largest $u$-$v$ spacings ($\ge$ 0.4 k$\lambda$). By doing so, they filtered 
out emission on large scales. As they noted, it is an efficient approach 
to get rid of most of the confusing interstellar emission. However, 
excluding the short spacings from the C-configuration data removes also 
all signatures of the shell. In particular for \alphaori at $\sim$+3\,\kms, 
it removed the source of $\sim$\,4$'$ in size associated with \alphaori 
which was detected by the NRT. If the same filtering is applied to the merged 
C+D data, only the ring of emission peaks is left, as can be seen in 
the 3.7 \kms ~channel map of Fig.~\ref{VLA-channel-maps-HR}.

Conversely, when we isolate the data obtained in configuration D, we only 
detect unambiguously a point source around --8\,\kms. The channel maps at LSR 
velocities larger than --5\,\kms ~suffer heavily from interstellar emission. 
In spite of this confusion, a source is detected at +3\,\kms, but it does 
not show the complex structure visible in Fig.~\ref{VLA-channel-map-3p1} 
because of the lower spatial resolution. Also a peak of emission is visible 
around +17\,\kms, but it is immersed in a larger structure.  
The combination of data obtained in configurations C and D is thus essential 
to reveal the full range of features that have been described above. 

\section[]{Interpretation}
\label{interpretation}

A star which is undergoing mass loss gets progressively surrounded by an 
expanding circumstellar shell. However, the collision of the supersonic 
stellar wind with the surrounding medium (ISM or an older envelope) produces 
a shell of denser material that is slowly expanding (Lamers \& Cassinelli 
1999). ``Detached shells'' have been reported to be detected by IRAS around 
many evolved stars (Young et al. 1993a). The emission of the detached shells 
observed at 60 and 100 $\mu$m is commonly assumed to be continuum emission 
radiated thermally by dust (Kerschbaum et al. 2010). Young et al. (1993b)
developed a model of a dust shell expanding into the ISM that simulates 
the formation of a detached shell and accounts for the IRAS results. 

\begin{figure}
\centering
\epsfig{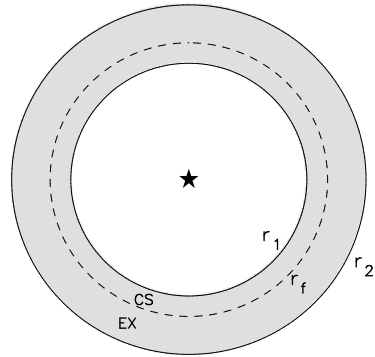}
\caption{Schematic view of the \alphaori detached shell. The termination shock 
is located in $r_1$, the contact discontinuity in $r_f$, and the bow shock 
in $r_2$. CS stands for circumstellar material and EX, for external 
material.}
  \label{schema}
\end{figure}

The Young et al. (1993b) scenario was revived by Libert et al. (2007) 
who developed a spherical model adapted to the gas and to the modeling of its 
\HI line emission. In this model a detached shell is produced by a stellar 
outflow that is abruptly slowed down at a termination shock. This first shock 
defines the inner limit of the detached shell (r$_1$). The outer limit (r$_2$) 
is defined by the leading shock (bow shock) where external matter is 
compressed by the expanding shell (Fig.~\ref{schema}). 
Between these two limits, the detached 
shell is composed of compressed circumstellar matter and interstellar 
matter separated by a contact discontinuity at r$_f$. The circumstellar matter 
is decelerated and heated when crossing the shock in r$_1$. As it cools down 
from T$_1^+$ (in r$_1$) to T$_f$ (in r$_f$), the expansion velocity decreases 
further, from v$_1^+$ to v$_f$, and the density increases from n$_1^+$ to 
n$_f^-$. Adopting an arbitrary temperature profile, Libert et al. (2007) 
solve numerically the equation of motion between r$_1$ and r$_f$. 
The temperature profile in the detached shell is constrained by the observed 
\HI line profiles. For the external part of the detached shell (between r$_f$ 
and r$_2$), Libert et al. (2007) assume a density profile that falls off as 
1/r$^2$ with the condition that the total mass (M$_{DT,CS}$) 
is given by the equivalent ISM mass enclosed in a sphere of radius r$_2$. 
Inside r$_1$, Libert et al. assume a spherical wind in uniform expansion. 

Originally Libert et al. (2007) developed this model for Y CVn, a carbon star 
surrounded by a dust detached shell discovered by IRAS (Young et al. 1993a) 
and imaged by ISO (Izumiura et al. 1996). \HI line profiles obtained at 
different positions on and around the central star were well fitted by such 
a model with 
a gas temperature decreasing from 1800\,K, in r$_1$, to 170 K, in r$_f$. 
Libert et al. applied the same model to other sources, like RX Lep, 
an oxygen-rich star (Libert et al. 2008), and V1942\,Sgr, 
another carbon star (Libert et al. 2010). 
In general, the gas detached shell model of Libert et al. (2007) 
accounts well for the \HI emission of evolved stars, which are characterized 
by a narrow line profile ($\sim$ 2-4 \kms) approximately centered on the 
stellar radial velocity (G\'erard \& Le~Bertre 2006). Indeed the emission 
from the detached shell where matter has been accumulating generally dominates 
the emission from the freely expanding wind region inside r$_1$. 
The \HI emission from an unresolved spherical wind would be characterized by 
a rectangular line profile of width twice the expansion velocity (Gardan 
et al. 2006). 

However, at high spatial resolution, the inner region ($<$\,r$_1$) may be 
resolved. In such a case a spectrum obtained on the stellar 
position should show two horns separated by twice the expansion velocity. 
In addition, a central line, corresponding to the detached shell should be 
observed in the middle, at the stellar radial velocity. Thus, for mass losing 
evolved stars, the Libert et al. (2007) model predicts a characteristic 
three-peaked profile at the central position, and a single-line profile away 
from it ($>$\,r$_1$) with a maximum intensity close to the inner shell 
boundary. 

\begin{figure}
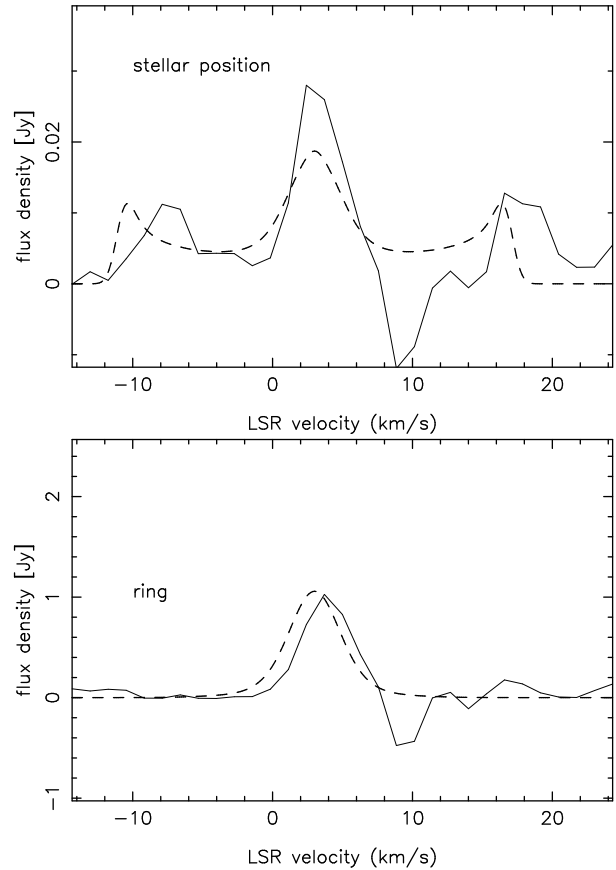

\centering
\epsfig{figure=newflux-centre-VLAplusmodel19-PUb.ps,angle=-90,width=8.0cm}
\epsfig{figure=newflux-120arcsec-VLAplusmodel19-PUb.ps,angle=-90,width=8.0cm}
\caption{Upper panel: \HI profile obtained on $\alpha$ Ori with the VLA 
within an aperture of 34$''$ diameter. 
Lower panel: \HI profile obtained by integrating the emission within a ring 
of internal radius, 80$''$, and external radius, 160$''$, and centered on 
the star. The dashed lines correspond to the modeled spectra 
discussed in Sect.~\ref{interpretation}.}
  \label{VLAplusmodel}
\end{figure}

The merged VLA data show two peaks of emission centered on \alphaori 
at \Vstar $\pm$ \Vexp. Close to \Vstar, one notes also a peak on the star 
(Fig.~\ref{VLA-channel-map-3p1}, upper panel) and a ring of emission 
around the star (Fig.~\ref{VLA-channel-map-3p1}, lower panel). 
Fig.~\ref{VLAplusmodel} shows spectra that have been extracted from the VLA 
merged data set at the position of the star and in a ring of internal radius, 
80$''$, and external radius, 160$''$. We expect that such a ring encloses 
most of the \HI emission from the detached shell. 
In the upper panel of Fig.~\ref{VLAplusmodel}, 
the predicted three-peaked profile is seen, with the peaks close to 
the expected velocities. On the other hand a single peak at \Vstar ~is 
visible in the spectrum shown in the lower panel. It thus seems that  
the VLA data contain information which, at least qualitatively, agrees 
with the Libert et al. (2007) model. However, the emission intrinsic to 
\alphaori is partly hidden by the confusion due to the ISM emission, as 
evidenced by the negative peaks at 9\,\kms. 
Similarly, the NRT data show an emission feature centered close to \Vstar. 
Due to its large beam, the NRT cannot detect the emission 
from the free-flowing wind of $\alpha$ Ori.

In view of these results, we have run the Libert et al. (2007) model 
with parameters (distance, expansion velocity, etc.) that apply 
to $\alpha$ Ori ~(see Sect.~\ref{basicdata}). 
As the stellar effective temperature is larger than 
2500\,K, we assume that all hydrogen is in atomic form (Glassgold \& Huggins 
1983). We assume also that 90\% of the mass is in hydrogen and 10\% in 
helium, so that the mean molecular weight, $\mu$, is equal to 1.3. 
An expansion velocity of the free wind, v$_0$\,=\,14\,\kms, implies a 
post-shock temperature in r$_1$ of almost 5800\,K (see Table~\ref{modelfit}). 
For the internal radius we select a value of 2$'$ ($\equiv$\,0.12 pc) 
in order to get a maximum \HI intensity at a position corresponding to 
the emission peaks identified in Fig.~\ref{VLA-channel-map-3p1}. 
The external radius was set 
arbitrarily to r$_2$=3$'$, and the ISM density to n$_{\rm H}$=1\,cm$^{-3}$. 
The calculated \HI emission is convolved with the telescope beam profile. 
For the NRT, we adopt the response of a rectangular aperture of 
160\,m$\times$30\,m (Gardan et al. 2006), and for the VLA a Gaussian response 
of FWHM\,=\,34$''$.

We have tried several combinations of mass-loss rate and duration of the mass 
loss episode. The mass-loss rate is constrained by the intensity of the peaks 
observed at \Vstar$\pm$\Vexp ~(i.e. --10 and +18 \kms) on the star. The mass 
in atomic hydrogen in the detached shell (which can be estimated by the 
emission feature observed at \Vstar) constrains the product of the mass loss 
rate by the duration. We found satisfactory fits for values of \.M in the 
range 1.0 to 1.5$\times$10$^{-6}$ \Msold ~and values of the duration 
in the range 0.7 to 1.0$\times$10$^5$ years. 

In Figs.~\ref{NRTplusmodel} and \ref{VLAplusmodel} we present the results 
of our best compromise that was obtained with \.M = 1.2$\times$10$^{-6}$ 
\Msold ~and a total duration of 0.8$\times$10$^5$ years. 
The parameters corresponding to this preferred model are summarized in 
Table~\ref{modelfit}. 
A rapid cooling of the gas (with a temperature assumed to vary 
as a power law from 5800\,K in r$_1$ to 220\,K in r$_f$) is 
needed in order to reproduce the observed narrow line width ($\sim$ 3 \kms) 
of the emission coming from the detached shell. 

In this spherical modeling we have adopted a systemic radial velocity 
of 3 \kms, which gives the best agreement with NRT data. The \HI emission 
from AGB stars is known to be sometimes shifted by $\sim$1--3 \kms ~with 
respect to the CO emission which is believed to provide the most precise 
estimate of the stellar radial velocity (G\'erard \& Le~Bertre 2006). 
In general this shift is towards 0 \kms ~(LSR), probably an effect of the 
dragging of the circumstellar shell by the ambient ISM. 
In Fig.~\ref{VLAplusmodel} (upper panel) it is clear that the central peak 
is not centered at the average of the two horn velocities. 
Consistently the model predicts horns that are blueshifted 
by $\sim$\,1\,\kms.
 
\begin{table}
\centering
\caption{Model parameters (d = 200 pc). The notations are as in Libert et al. 
(2007). In particular, t$_{DS}$ is the formation time of the detached shell, 
M$_{DT,CS}$ is the mass of the circumstellar component of the detached shell, 
and M$_{DT,EX}$ is the external mass accreted in the detached shell.}
\begin{tabular}{ll}
\hline
\.M                               & 1.2$\times$10$^{-6}$ \Msold\\
$\mu$                             & 1.3\\
t$_1$                             & 8\,140 years\\
t$_{DS}$                          & 71\,860 years\\
r$_1$                             & 0.12 pc (2.0$'$)\\
r$_f$                             & 0.14 pc (2.35$'$)\\
r$_2$                             & 0.18 pc (3.0$'$)\\
T$_0$($\equiv$ T$_1^-$), T$_1^+$  & 20 K, 5796 K\\
T$_f$ (= T$_2$)                   & 223 K\\
v$_0$($\equiv$ v$_1^-$), v$_1^+$  & 14 \kms, 3.51 \kms\\
v$_f$                             & 0.06 \kms\\
v$_2$                             & 0.9 \kms\\
n$_1^-$, n$_1^+$                  & 14.0 H\,cm$^{-3}$, 55.8 H\,cm$^{-3}$\\
n$_f^-$, n$_f^+$       & 1.8$\times$10$^{3}$ H\,cm$^{-3}$, 2.5 H\,cm$^{-3}$\\
n$_2$                             & 1.6 H\,cm$^{-3}$\\
M$_{r < r_1}$                     & 9.8$\times$10$^{-3}$ \Msol\\
M$_{DT,CS}$                       & 0.086 \Msol\\
M$_{DT,EX}$                       & 0.8$\times$10$^{-3}$ \Msol\\
\hline
\end{tabular}
\label{modelfit}
\end{table}

\section[]{Discussion}
\label{discussion}

With the NRT and the VLA, we have detected an \HI source coincident with 
\alphaori and relatively compact ($\sim 4'$ in right ascension). The emission 
is centered at \Vlsr $\sim$ 3 \kms, ~and has a narrow line profile (FWHM 
$\sim$ 3 \kms). In addition, the VLA channel map at 3.7 \kms ~shows a ring of 
emission encircling the position of the star and a tail extending south-west. 
Furthermore, the VLA data show emission lines at $-$9\,\kms ~and at +18 \kms, 
right at the position of the central star. Although our data are affected by 
confusion, these features can be clearly ascribed to the circumstellar shell 
expelled by the supergiant.

These features can be explained qualitatively by the Libert et al. (2007) 
detached shell model. In particular, the spectrally narrow emission observed 
close to \Vstar, which the VLA shows to peak along a ring encircling 
the position of \alphaori at 3.7 \kms, 
is characteristic of the emission expected from a detached shell. 
Using parameters which are representative of $\alpha$ Ori, the model 
gives a fair agreement with the observations. The main result is that the 
detached gas shell of \alphaori can be accounted for by a mass loss episode 
of $\sim$\,10$^5$ years duration at a rate in the range 1-2$\times$10$^{-6}$ 
\Msold. 

In Sect.~\ref{VLAobsCplusD}, we noted that Bowers \& Knapp (1987) detected 
only the horns at \Vstar$\pm$\Vexp ~corresponding to the free flowing wind 
of $\alpha$ Ori, because they had selected large $u$-$v$ spacings. 
They modeled these horns with a wind of 2.2$\times$10$^{-6}$ \Msold 
~(at 200 pc). In fact this agrees quite well with our model, because 
they adopted an envelope radius of 60$''$, half as much as our own r$_1$ 
estimate. The reprocessing of their full data set and the combination 
with D-configuration data revealed the stationary shell of size $\sim$4$'$ 
that they missed. 

Here we want to stress again the importance of multi-scale observations, 
and in particular of the C-configuration data that 
we have retrieved from the VLA archive. Indeed, it is the improved angular 
resolution that allows us to characterize geometrically the detached shell. 
For three other sources that we have studied with the VLA in configuration 
D, Mira (Matthews et al. 2008), RS~Cnc (Matthews \& Reid 2007, Libert et al. 
2010) and X~Her (Matthews et al. 2011), we have not observed shells that are 
seen to be detached from the central stars. However the observed \HI 
line-profiles indicate that the stellar winds are slowed down by the external 
medium, and that the outer circumstellar shells should be quasi-stationary 
with an enhanced density. Thus we expect that observations of these sources 
with a higher spatial resolution would reveal a detachment of the outer 
shells.

The mass-loss rate that we find is smaller than the mass-loss rate used by 
Meynet \& Maeder (2003) in the evolution of a 20-\Msol ~star which, among 
their models, gives the best agreement with the temperature and luminosity of 
$\alpha$ Ori. A better agreement on the temperature and 
mass-loss rate might be found for models with a lower initial mass of 
$\alpha$ Ori, but also with a lower luminosity, which would imply 
a distance lower than 150\,pc. We note however that, for the mass loss 
of cool stars, Meynet \& Maeder (2003) use prescriptions 
from de Jager et al. (1988), which produce in their models smooth variations 
of the mass-loss rate with time. 
This is only indicative, as mass loss is known to be strongly variable in the 
progenitors of red supergiants, especially when they cross the instability 
strip. 

From observations, the mass-loss rate of \alphaori is uncertain, and values 
ranging over a factor 10 have appeared in the literature. For instance 
De Beck et al. (2010), from a modeling of several CO rotational lines, derive 
a mass loss rate of $\sim$ 5 $\times$ 10$^{-7}$ \Msold ~(for a distance 
of 200 pc), whereas Harper et al. (2001, 2009), from a model of the radio 
continuum emission, obtained 4.8~($\pm$1.3)$\times$ 10$^{-6}$ \Msold 
~(for the same distance). This large range arises in part from tracers 
probing different regions of the circumstellar shell, and from a rate of mass 
loss that has likely been variable. Also the CO emission seems anomalously 
low in red supergiants which could lead to an underestimate of the 
mass-loss rate (Josselin et al. 1998). 
Interferometric data obtained at 11 $\mu$m by Danchi et al. (1994) seem to 
show the presence of several dust shells close (at 1 and 2$''$) to the central 
star, suggesting a mass loss rate that is highly variable. We have adopted 
a value (1.2 $\times$ 10$^{-6}$ \Msold) which, in our model, represents 
an average over 8000 years.

In our modeling the gas temperature decreases from 5800 K 
(in r$_1$) to 220 K (in r$_f$) in a lapse of 72$\times$10$^3$ years.  
This translates to an average cooling time-scale of ~ 2.2$\times$10$^4$ years. 
The post-shock density is 56\,H\,cm$^{-3}$, and rises up to 1800\,H\,cm$^{-3}$
at the contact discontinuity. Spitzer (1968) gives a cooling time of about 
2.5$\times$10$^4$ years for a gas of density 10\,H\,cm$^{-3}$ and temperature 
1000 K. Therefore the modeling is consistent with a quite reasonable 
hypothesis for the cooling. The cooling per unit mass is proportional to   
the density. From Dalgarno \& McCray (1972) it should also depend on 
the electron density and on the temperature. 
From our modelling we can estimate the temperature and the density 
as a function of the distance to the central star. 
The electron density is an unknown. However, hydrogen should stay atomic 
within the detached shell, but species with a low ionization potential, 
such as carbon, are expected to be singly ionized (Huggins et al. 1994). 

\begin{figure}
\centering
\epsfig{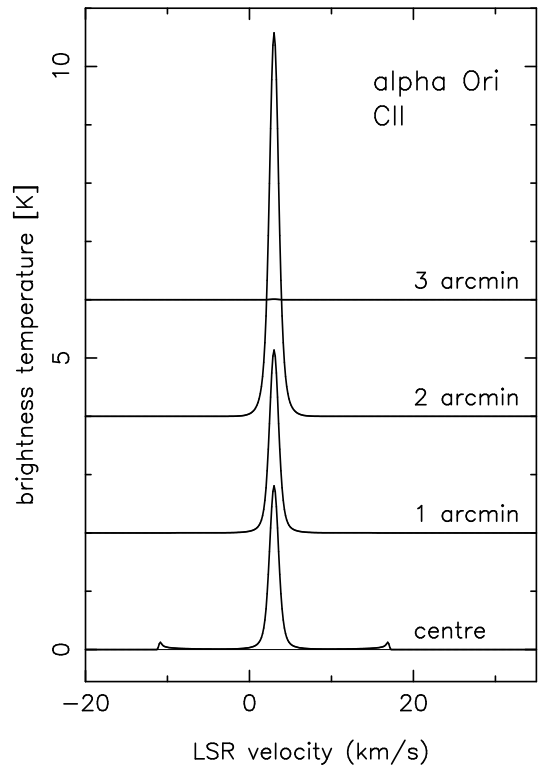}
\caption{Simulation of the \CII line emission at various distances from 
$\alpha$ Ori. For clarity, the spectra are successively shifted by 2\,K.}
  \label{simulCII}
\end{figure}

The cooling lines from the detached shell are predicted to be centered 
at the same velocity as the \HI line (3\,\kms), and to have a similarly 
narrow line-profile ($\sim$\,3\,\kms). As an example we give the profile 
of the \CII line at 158 $\mu$m at various positions across the detached shell 
(Fig.~\ref{simulCII}). We assumed carbon to have an abundance of 
2$\times$10$^{-4}$ relative to hydrogen and to be completely ionized 
beyond 6$\times10^{-3}$ pc (0.1$'$) from the central star. 
The calculation has been done assuming a gaussian beam of 12$''$  (FWHM),
and adopting the same temperature profile 
as for hydrogen (Sect.~\ref{interpretation}).  
This cooling line should be within reach of the IR Heterodyne Spectrometer 
on the SOFIA (Stratospheric Observatory for Infrared Astronomy). 

The efficient cooling that we 
have adopted implies that the detached gas shell in the model is geometrically 
narrow (r$_f$/r$_1$ $\sim$ 1.2). This could be an explanation for the relative 
compactness of the \HI source ($\phi \sim 4'$) and for the squeezing of the 
contours around the central star (except south-west) in the 2.4 and 3.7 \kms 
~channel maps (Figs.~\ref{VLA-channel-map-3p1} and 
\ref{VLA-channel-map-3p1-restricted}). 

The model that we are using is spherical and cannot account for all the 
features observed in \alphaori, which is known to be moving through the 
ISM at $\sim$\,30\,\kms ~towards the north-east. The plateau that is observed 
south-west in the 3.7 \kms ~channel map could be the signature of a tail 
of material such as that observed behind Mira (Matthews et al. 2008). 
This kind of feature is predicted by 3D hydrodynamic simulations of CSEs 
around mass losing stars moving through the ISM, such as those performed 
by Villaver et al. (2003) or by Wareing et al. (2007), and might hopefully 
contain information on past episodes of mass loss. 

The \HI emission that we observe in \alphaori is concentrated in a source of 
diameter $\sim\,4'$, and is thus found well inside the infrared emission 
reported by Noriega-Crespo et al. (1997) and by Ueta et al. (2008), and that 
Herschel has shown to arise from several thin shells of radii $\sim$\,6$'$ 
(Cox et al. 2012). These shells are clearly distinct from the classical 
detached shells of Young et al. (1993a). The mechanism responsible for 
the far-infrared 
emission of these shells has not been identified with certainty. 
It could be emission by dust 
heated by collisions with the shocked gas and/or emission from forbidden lines 
of O\,{\sc {i}} and C\,{\sc {ii}}. However there is a consensus that 
this emission traces the bow shock(s) resulting from the motion of \alphaori 
through the ISM. van Marle et al. (2011) have studied the coupling of dust 
and gas within a shocked wind with conditions typical of $\alpha$\,Ori. 
They find that, apart for large grains (radius~$>$~0.045\,$\mu$m), 
the dust remains bound to the gas in such a structure. 

We found however that the far-IR arc coincides with FUV emission and 
is associated with atomic hydrogen emission (Figs.~\ref{VLA-channel-maps-HR} 
and \ref{VLA-channel-maps-plus-IRAS}). The mass in atomic hydrogen compares 
fairly well with the estimate of the total mass of the far-IR arc by 
Noriega-Crespo et al. (1997), given the uncertainties in our \HI estimate and 
in the gas-to-dust ratio (100) adopted by Noriega-Crespo et al. \HI emission 
is found mainly at 7.6$\pm$1.3 \kms, i.e. at the velocity at which 
interstellar emission peaks (see Fig.~\ref{NRT-LAB}). Thus the atomic gas 
that we detect in coincidence with the far-IR arc seems related to the 
surrounding ISM rather than to $\alpha$ Ori. Possibly, it is interstellar 
matter compressed in a precursor ahead of the bow shock.

The ISM gas that crosses the bow shock at $\sim$\,30\,\kms ~should be heated 
at a temperature $\sim$\,25\,000\,K, and hydrogen may be ionized. 
Except to the south-west, we have found no unambiguous evidence of atomic 
hydrogen associated with \alphaori beyond 3$'$ from the star. This suggests 
that hydrogen in the region between the interface and the bow shock remains 
ionized during the lapse of $\sim$\,10$^5$ years that we infer 
for the formation of the detached shell, or that it remains sufficiently warm 
that its emission is widened over a large range of velocities, preventing 
its detection. In these conditions, our \HI 
modeling cannot bring real constraints on the region of the detached shell 
which is filled by matter from the ISM. In particular, the value that we 
have adopted for the parameter r$_2$ has no real significance. 

Conversely, 
the parameters that we find for the circumstellar part of the detached shell 
could be used as constraints for the 3D hydrodynamic simulations 
of the Betelgeuse's bow shock that start to be developed (e.g. Mohamed et al. 
2011). In this specific case, we note that the interface is at \gsim\,0.3\,pc 
from the central star, which seems too large compared to our \HI 
observations, that suggest a smaller value. 
Possibly the rate of mass loss adopted by Mohamed et al. 
(3\,$\times$\,10$^{-6}$\,\Msold) is too large. The density of the ISM 
material through which \alphaori is moving is also uncertain. 

Another possibility is that the far-IR arc seen at 6$'$ from \alphaori ~is not 
the true bow shock related to the detached gas shell that is detected at 
21\,cm. The examination of the images published by Cox et al. (2012) suggests 
that there are other structures inside the far-IR arc, in particular 
another arc at 3$'$ north-east of $\alpha$ Ori. It is tempting to hypothesize 
that this 3$'$-radius arc is the true bow shock related to the detached shell 
that we have modeled. In that case the 6$'$-radius arc which is also seen in 
\HI and far-UV would be related to a more ancient episode of mass loss. 
Clearly spectral information on the far-IR emission of this arc and imaging 
in \HI at a higher spatial resolution are needed in order to elaborate on its 
true nature. The characteristics of the internal medium, between the detached 
shell and the far-IR arc, are also an issue. 

\section{Conclusions}

Observations in the \HI line at 21 cm of \alphaori have been performed with 
the NRT and the VLA. A source ($\phi \sim 4'$) has been detected at 
the position of the star. The emission is dominated by a narrow line (FWHM 
$\sim$ 3 \kms) centered close to \Vstar. The VLA data show that this 
source has the structure of a detached shell elongated towards the south-west. 
This shape agrees with the numerical simulations of a CSE moving through the 
ISM. Emission peaks at \Vstar$\pm$\Vexp ~are also observed by the VLA at 
the position of the central star. Presently the data bring no clear evidence 
of atomic hydrogen outside this source. 

We have developed a working model adapted to \alphaori and its detached 
neutral gas  shell that accounts for the spectral features that are observed 
and for the size of the source. In this model, the detached gas shell 
of \alphaori results from the collision of a 1.2$\times$10$^{-6}$ \Msold 
~stellar wind, expanding at 14\,\kms, with the surrounding ISM for a lapse 
of $\sim$\,10$^5$ years. The gas is cooling from a post-termination-shock 
temperature of 5800\,K to $\sim$220\,K, at the interface with ISM.

The high level of confusion in the direction of $\alpha$ Ori forces 
us to be cautious. Nevertheless, by selecting different ranges of baselines, 
it is possible to obtain images revealing the detached shell of $\alpha$ Ori 
as well as \HI emission coincident with the 6$'$ far-IR/far-UV arc. 
Presently, we cannot conclude on the exact nature of the structure in arc 
discovered by IRAS. We suggest that the \HI emission might come from  
a precursor of the \alphaori bow shock, or that the whole structure could 
be related to an ancient episode of mass loss.

\section*{Acknowledgments}
The Nan\c{c}ay Radio Observatory is the Unit\'e scientifique de Nan\c{c}ay of 
the Observatoire de Paris, associated as Unit\'e de Service et de Recherche 
(USR) No. B704 to the French Centre National de la Recherche Scientifique 
(CNRS). The Nan\c{c}ay Observatory also gratefully acknowledges the financial 
support of the Conseil R\'egional de la R\'egion Centre in France. 
The VLA observations presented here are part of the NRAO program AM1001. 
LDM acknowledges support from grant AST-1009644 from the National Science 
Foundation. This research has made use of the SIMBAD and ADS databases. 
We thank Dr. Graham Harper for his careful reading of the manuscript and 
useful comments.

\label{lastpage}


\begin{thebibliography}{}
\bibitem[Bowers \& Knapp(1987)]{bk87}
Bowers, P. F., \& Knapp, G. R., 1987, ApJ, 315, 305 
\bibitem[Bowers \& Knapp(1988)]{bk88}
Bowers, P. F., \& Knapp, G. R., 1988, ApJ, 332, 299 
\bibitem[Chiavassa et al. (2011)]{chiav11}
Chiavassa, A., Haubois, X., Young, J. S., Plez, B., Josselin, E., 
Perrin, G., \&  Freytag, B., 2011, A\&A, 515, A12 
\bibitem[Cox et al. (2012)]{cox12}
Cox, N. L. J., Kerschbaum, F., van Marle, A.-J., et al., 2012, A\&A, 537, A35 
\bibitem[Dalgarno \& McCray (1972)]{dm72}
Dalgarno, A., \& McCray, R. A., 1972, ARAA, 10, 375
\bibitem[Danchi et al. (1994)]{danchi94}
Danchi, W. C., Bester, M., Degiacomi, C. G., Greenhill, L. J., \& 
Townes, C. H., 1994, AJ, 107, 1469
\bibitem[De Beck et al. (2010)]{debeck2010}
De Beck, E., Decin, L., de Koter, A., et al., 2010, A\&A, 523, A18
\bibitem[de Jager et al. (1988)]{dejager88}
de Jager, C., Nieuwenhuijzen, H., \& van der Hucht, K. A., 1988, A\&AS, 72, 259
\bibitem[Gardan et al. (2006)]{gardan06}
Gardan, E., G\'erard, E., \& Le~Bertre, T., 2006, MNRAS, 365, 245
\bibitem[G\'erard \& Le~Bertre (2006)]{glb06}
G\'erard, E., \& Le~Bertre, T., 2006, AJ, 132, 2566 
\bibitem[G\'erard et al. (2011)]{gerard11}
G\'erard, E., Le~Bertre, T., \& Libert, Y., 2011, Proc. ``{\it SF2A 2011''}, 
G. Alecian, K. Belkacem, S. Collin, R. Samadi \& D. Valls-Gabaud (eds.), 
p.~419 
\bibitem[Glassgold \& Huggins (1983)]{gh83}
Glassgold, A. E., \& Huggins, P. J., 1983, MNRAS, 203, 517
\bibitem[Harper et al. (2008)]{harper08}
Harper, G. M., Brown, A., \& Guinan, E. F., 2008, AJ, 135, 1430
\bibitem[Harper et al. (2001)]{harper01}
Harper, G. M., Brown, A., \& Lim, J., 2001, ApJ, 551, 1073 
\bibitem[Harper et al. (2009)]{harper09}
Harper, G. M., Richter, M. J., Ryde, N., Brown, A., Brown, J., Greathouse, 
T. K., \& Strong, S., 2009, ApJ, 701, 1464
\bibitem[Haubois et al. (2009]{hautbois09}
Haubois, X., Perrin, G., Lacour, S., et al., 2009, A\&A, 508, 923
\bibitem[Huggins (1987)]{huggins87}
Huggins, P. J., 1987, ApJ, 313, 400
\bibitem[Huggins et al. (1994)]{huggins94}
Huggins, P. J., Bachiller, R., Cox, P., \& Forveille, T., 1994, ApJ, 424, L127
\bibitem[Izumiura et al. (1996)]{izumiura96}
Izumiura, H., Hashimoto, O., Kawara, K., Yamamura, I., \& Waters, L. B. F. M.,
1996, A\&A, 315, L221
\bibitem[Josselin et al. (1998)]{josselin98}
Josselin, E., Loup, C., Omont, A., Barnbaum, C., Nyman, L.-\AA., \& 
S\`evre, F., 1998, A\&AS, 129, 45
\bibitem[Josselin \& Plez (2007)]{jp07}
Josselin, E., \& Plez, B., 2007, A\&A, 469, 671
\bibitem[Kalberla et al. (2005)]{kalberla05}
Kalberla, P. M. W., Burton, W. B., Hartmann, D., Arnal, E. M., Bajaja, E., 
Morras, R., \& P\"oppel, W. G. L., 2005, A\&A, 440, 775
\bibitem[Kerschbaum et al. (2010)]{kerschbaum10}
Kerschbaum, F., Ladjal, D., Ottensamer, R., et al., 2010, A\&A, 518, L140
\bibitem[Kervella et al. (2011)]{kervella11}
Kervella, P., Perrin, G., Chiavassa, A., Ridgway, S. T., Cami, J., 
Haubois, X., \& Verhoelst, T., 2011, A\&A, 531, A117
\bibitem[Lamers \& Cassinelli (1999)]{lc99}
Lamers, J. G. L. M., \& Cassinelli, J. P., 1999, ``Introduction to Stellar 
Winds'', Cambridge University Press, Chap.~12
\bibitem[Levesque et al. (2005)]{levesque05}
Levesque, E. M., Massey, P., Olsen, K. A. G., et al., 2005, ApJ, 628, 973
\bibitem[Libert et al. (2007)]{lib07}
Libert, Y., G\'erard, E., \& Le~Bertre, T., 2007, MNRAS, 380, 1161
\bibitem[Libert et al. (2010)]{lib10a}
Libert, Y., G\'erard, E., Thum, C., Winters, J.~M., Matthews, L. D., \& 
Le~Bertre, T., 2010, A\&A, 510, A14 
\bibitem[Libert et al. (2008)]{lib08}
Libert, Y., Le\,Bertre, T., G\'erard, E., \&  Winters, J.\,M., 2008, A\&A, 
491, 789 
\bibitem[Libert et al. (2010)]{lib10b}
Libert, Y., Winters, J.\,M., Le\,Bertre, T., G\'erard, E., \& Matthews, L. D., 
2010, A\&A, 515, A112
\bibitem[Lim et al. (1998)]{lim98}
Lim, J., Carilli, C. L., White, S. M., Beasley, A. J., \& Marson, R. G., 1998, 
Nature, 392, 575
\bibitem[Martin et al. (2005)]{martin05}
Martin, D. C.,  Fanson, J., Schiminovich, D., et al., 2005, ApJ, 619, L1
\bibitem[Matthews et al. (2011)]{matthews11}
Matthews, L. D., Libert, Y., G\'erard, E., Le\,Bertre, T., Johnson, M. C., 
Dame, T. M., 2011, AJ, 141, id. 60
\bibitem[Matthews et al. (2008)]{matthews08}
Matthews, L. D., Libert, Y., G\'erard, E., Le\,Bertre, T., \& Reid, M. J.,  
2008, ApJ, 684, 603 
\bibitem[Matthews \& Reid (2007)]{mr07}
Matthews, L. D., \& Reid, M. J., 2007, AJ, 133, 2291
\bibitem[Meynet \& Maeder (2003)]{mm03}
Meynet, G., \& Maeder, A., 2003, A\&A, 404, 975
\bibitem[Mohamed et al. (2011)]{mohamed11}
Mohamed, S., Mackey, J., \& Langer, N., 2011, A\&A, in press (arXiv:1109.1555) 
\bibitem[Morrissey et al. (2007)]{morrissey07}
Morrissey, P., Conrow, T., Barlow, T. A., et al., 2007, ApJS, 173, 682
\bibitem[Neilson et al. (2011)]{neilson11}
Neilson, H. R., Lester, J. B., \& Haubois, X., 2011, to appear in the 
proceedings of "The 9$^{\rm th}$  Pacific Rim Conference on Stellar 
Astrophysics", ASP Conf. Ser. (arXiv1109.4562)
\bibitem[Noriega-Crespo et al. (1997)]{nc97}
Noriega-Crespo, A., van Buren, D., Cao, Y., \& Dgani, R., 1997, AJ, 114, 837
\bibitem[Ohnaka et al. (2011)]{ohnaka11}
Ohnaka, K., Weigelt, G., Millour, F., et al., 2011, A\&A, 529, A163
\bibitem[Perrin et al. (2004)]{perrin04}
Perrin, G., Ridgway, S. T., Coud\'e du Foresto, V., Mennesson, B., Traub,  
W. A., \& Lacasse, M. G., 2004, A\&A, 418, 675
\bibitem[Perryman et al. (1997)]{perryman97}
Perryman, M. A. C., Lindegren, L., Kovalevsky, J., et al., 1997, A\&A, 323, 
L49
\bibitem[Sahai \& Chronopoulos (2010)]{sc10}
Sahai, R., \& Chronopoulos, C. K., 2010, ApJ, 711, L53 
\bibitem[Sch{\"o}nrich et al. (2010)]{schonrich10}
Sch{\"o}nrich, R., Binney, J., \& Dehnen, W., 2010, MNRAS, 403, 1829
\bibitem[Spitzer (1968)]{spitzer68}
Spitzer, L., 1968, "Diffuse Matter in Space", Interscience Publishers, p.~139
\bibitem[Ueta et al. (2008)]{ueta08}
Ueta, T., Izumiura, H., Yamamura, I., et al., 2008, PASJ, 60, S407
\bibitem[van Leeuwen (2007)]{vl07}
van Leeuwen, F., 2007, "Hipparcos, the New Reduction of the Raw Data", 
Springer, Astrophysics and Space Science Library, vol. 350
\bibitem[van Loon (2010)]{vl10}
van Loon, J. T., 2010, ASP Conf. Ser., 425, 279
\bibitem[van Marle et al. (2011)]{vm11}
van Marle, A. J., Meliani, Z., Keppens, R., \& Decin, L., 2011, ApJ, 734, L26
\bibitem[Villaver et al. (2003)]{villaver03}
Villaver, E., Garc\'ia-Segura, G., \& Manchado, A., 2003, ApJ, 585, L49
\bibitem[Wareing et al. (2007)]{wareing07}
Wareing, C. J., Zijlstra, A. A., \& O'Brien, T. J., 2007, MNRAS, 382, 1233
\bibitem[Young et al. (1993a)]{young93a}
Young, K., Phillips, T. G., \& Knapp, G. R., 1993a, ApJS, 86, 517
\bibitem[Young et al. (1993b)]{young93b}
Young, K., Phillips, T. G., \& Knapp, G. R., 1993b, ApJ, 409, 725
\end{thebibliography}
\end{document}